\documentclass[]{interact}
\usepackage{epstopdf}% To incorporate .eps illustrations using PDFLaTeX, etc.
\usepackage{subfigure}% Support for small, `sub' figures and tables
\usepackage{url}
\usepackage{xcolor}
\usepackage{threeparttable}
\usepackage[numbers,sort&compress]{natbib}% Citation support using natbib.sty
\bibpunct[, ]{[}{]}{,}{n}{,}{,}% Citation support using natbib.sty
\usepackage{amsmath}
\theoremstyle{plain}% Theorem-like structures

\theoremstyle{definition}

\usepackage{multirow}
\theoremstyle{remark}

\usepackage{amssymb}
\usepackage{amsthm}

\begin{document}

\articletype{Communications in Materials Informatics}

%\title{ChemTS: Molecule Design using Monte Carlo Tree Search with Neural Rollout}
\title{ChemTS: An Efficient Python Library for {\em de novo} Molecular Generation}

\author{
\name{Xiufeng Yang\textsuperscript{a}\thanks{CONTACT Koji Tsuda. Email: tsuda@k.u-tokyo.ac.jp}, Jinzhe Zhang\textsuperscript{d}, Kazuki Yoshizoe\textsuperscript{c}, Kei Terayama\textsuperscript{a}, and Koji Tsuda\textsuperscript{a,b,c}}
\affil{\textsuperscript{a}Graduate School of Frontier Sciences, 
The University of Tokyo, 5-1-5 Kashiwanoha, Kashiwa, Chiba 277-8561, Japan;
\textsuperscript{b}National Institute for Materials Science,
1-2-1 Sengen, Tsukuba, Ibaraki 305-0047 Japan;
\textsuperscript{c}RIKEN, Center for Advanced Intelligence Project, 1-4-1
Nihombashi, Chuo, Tokyo 103-0027, Japan; \textsuperscript{d}Department
of Biosciences, INSA Lyon, 43 bd du 11 novembre 1918, 69622,
Villeurbanne Cedex, France}
}

\maketitle

\begin{abstract}
Automatic design of organic materials requires black-box optimization in a vast
chemical space. 
In conventional molecular design algorithms, 
a molecule is built as a combination of predetermined fragments. 
Recently, deep neural network models such as variational auto encoders (VAEs) 
and recurrent neural networks (RNNs) are shown to
be effective in {\em de novo} design of molecules 
without any predetermined fragments.
This paper presents a novel python library ChemTS that explores the chemical
space by combining Monte Carlo tree search (MCTS) and an RNN. 
In a benchmarking problem of optimizing the octanol-water partition
coefficient and synthesizability, our algorithm showed superior
efficiency in finding high-scoring molecules.  
ChemTS is available at \url{https://github.com/tsudalab/ChemTS}.
\end{abstract}


\section{Introduction}
In modern society, a variety of organic molecules 
are used as important materials such as solar cells~\cite{niu2015review}, 
organic light-emitting diodes (OLEDs)~\cite{kaji2015purely}, 
conductors~\cite{ueda2014hydrogen},
sensors~\cite{yeung2015luminescent}
and ferroelectrics~\cite{horiuchi2008organic}.
At the highest level of abstraction, design of organic molecules 
is formulated as a combinatorial optimization problem 
to find the best solutions in a vast chemical space.
Most computer-aided methods for molecular design build a molecule 
by a combination of predefined fragments
(e.g.,\cite{podlewska2017creating}).
Recently, Ikebata et al~\cite{ikebata2017bayesian} succeeded 
{\em de novo} molecular design using an engineered language model of
SMILES representation of molecules~\cite{weininger1988smiles}.
It is increasingly evident, however, that engineered models often
perform worse than neural networks in text and image 
generation~\cite{bowman2015generating,oord2016pixel}. 
Gomez-Bombarelli et al.~\cite{gomez2016automatic} were the first to 
employ a neural network called variational autoencoder (VAE) 
to generate molecules. 
Later Kusner et al. enhanced it to grammar variational 
autoencoder (GVAE)~\cite{kusner2017grammar}. 
SMILES strings created by VAEs are mostly invalid 
(i.e., they do not translate to chemical structures), 
so generation steps have to be repeated many times to obtain a molecule.
Segler et al.~\cite{segler2017generating} 
showed that a recurrent neural network (RNN) using long short term
memory (LSTM)~\cite{DBLP:conf/emnlp/ChoMGBBSB14} 
achieves a high probablity of 
valid SMILES generation. In their algorithm, a large number of
candidates are generated randomly and a black-box optimization algorithm
is employed to choose high-scoring molecules. 
It is required to generate a very large number of candidates 
to ensure that desirable molecules are included in the candidate set. 
Optimization in a too large candidate space can be inhibitively slow. 

In this paper, 
we present a novel python library ChemTS to 
offer material scientists a versatile tool of 
{\em de novo} molecular design.  
The space of SMILES strings is represented as a search tree where the
$i$-th level corresponds to the $i$-th symbol. 
A path from the root to a terminal node corresponds to a complete SMILES string.
Initially, only the root node exists and 
the search tree is gradually generated 
by Monte Carlo tree search (MCTS)~\cite{Browne2012}.
MCTS is a randomized best-first search method 
that showed exceptional performance in 
computer Go~\cite{silver2016mastering}. 
Recently, it has been successfully applied to alloy design~\cite{m2017mdts}. 
MCTS constructs only a shallow tree and downstream paths 
are generated by a {\rm rollout procedure}.
In ChemTS, an RNN trained by a large database of SMILES strings 
is used as the rollout procedure. 
In a benchmarking experiment, ChemTS showed better efficiency in
comparison to VAEs, creating about 40 molecules per minute.
As a result, high scoring molecules were generated within several hours.

%\begin{figure}[h!]
% \centering
%       \includegraphics[width=0.5\textwidth]{train.pdf}
%  \caption{Illustration of training the GRU. The GRU network is unrolled for $t$ time steps.}
%\label{rnn_train}
%\end{figure}

\section{Method}
ChemTS requires a database of SMILES strings and a reward 
function $r(S)$ where $S=\{s_1,\ldots,s_T\}$ is an input SMILES string.
Our definition of SMILES strings contains the following symbols 
representing atoms, bonds, ring numbers and branches: 
$s_t \in $ \{C, c, o, O, N,  F, [C@@H], n, -,  S,Cl, [O-],[C@H],
[NH+],[C@], s, Br,  [nH], [NH3+],  [NH2+], [C@@], [N+], [nH+], [S@],
[N-], [n+],[S@@], [S-], I, [n-], P, [OH+],[NH-], [P@@H], [P@@], [PH2], 
[P@], [P+], [S+],[o+], [CH2-], [CH-], [SH+], [O+], [s+], [PH+], [PH], 
[S@@+], /,=, \#, 1,2,3,4,5,6,7,8,9,(, ),\}. 
In addition, we have a terminal symbol \$.
The reward function involves first principle or semi-empirical calculations
and describes the quality of the molecule described by $S$.
If $S$ does not correspond to a valid molecule, $r(S)$ is set to an
exceptionally small value. 
We employ {\em rdkit} (\url{www.rdkit.org}) to check if $S$ is valid or not.
Before starting the search, an RNN is trained by the database and 
we obtain the conditional probability $P(s_t|s_1,...,s_{t-1})$ as a result.
The architecture of our RNN is similar to that in 
\cite{segler2017generating} and will be detailed in Section~\ref{sec:rnn}.

\begin{figure*}
  \includegraphics[width=\textwidth]{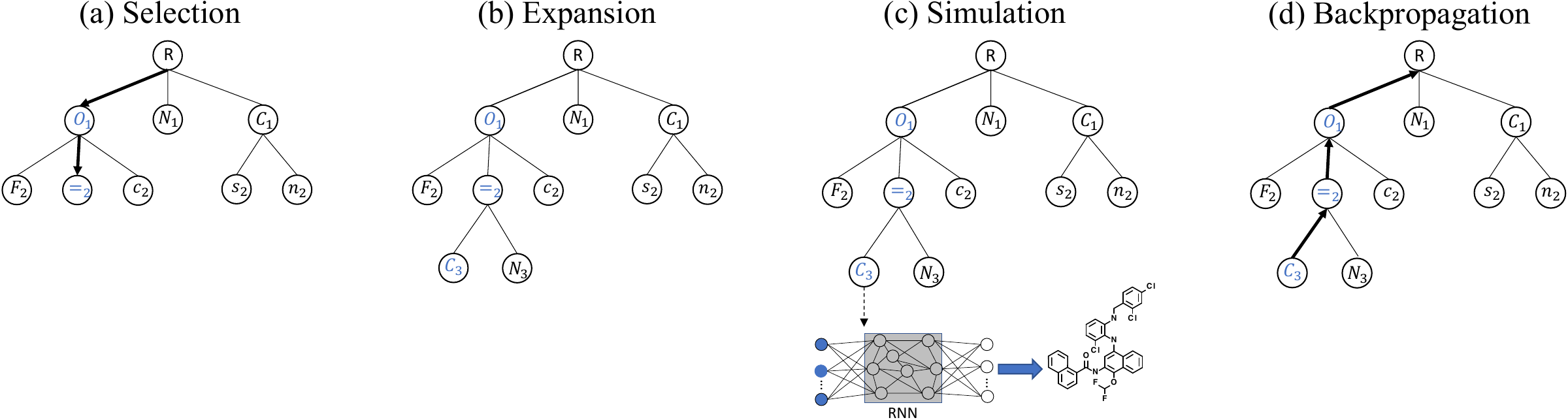}
  \caption{Monte Carlo Tree Search. (a) Selection step. The search tree
    is traversed from the root to a leaf by choosing the child with the largest UCB score.
(b) Expansion step. Children nodes are created by sampling from RNN 30 times. 
(c) Simulation step. Paths to terminal nodes are created by the rollout
    procedure using RNN. Rewards of the corresponding molecules are computed.
(d) Backpropagation step. The internal parameters of upstream nodes are updated.}
  \label{fig:mcts}
\end{figure*}

MCTS creates a search tree, where each node corresponds to one
symbol. Nodes with the terminal symbol are called terminal nodes. 
Starting with the root node, 
the search tree grows gradually by repeating the four steps, 
Selection, Expansion, Simulation and Backpropagation
(Figure~\ref{fig:mcts}). 
Each intermediate node has a UCB score 
that evaluates the merit of the node~\cite{Browne2012}.
The distinct feature of MCTS is the use of {\em rollout} 
in the simulation step. 
Whenever a new node is added, paths from the node to terminal nodes 
are built by a random process.
In computer games, it is known that uniformly random rollout does not
perform well, and designing a better rollout procedure 
based on available knowledge is essential in achieving high 
performance~\cite{Browne2012}.
Our idea is to employ a trained RNN for rollout. 
A node at level $t-1$ has a partial SMILES string $s_1,\ldots,s_{t-1}$
corresponding to the path from the root to the node.
Given the partial string, RNN allows us to compute the distribution of
the next letter $s_t$.
Sampling from the distribution, the string is elongated by one.  
Elongation by RNN is repeated until the terminal symbol occurs.
After elongation is done, the reward of the generated string is computed. 
In the backpropagation step, the reward is propagated backwards and 
the UCB scores of traversed nodes are updated. 
See \cite{m2017mdts} for details about MCTS.

\subsection{Recurrent Neural Network} \label{sec:rnn}
Our recurrent neural network (RNN) has a non-deterministic output: 
an input string $s_1,\ldots,s_T$ is mapped to 
probability distributions of output symbols $P(y_1),\ldots,P(y_T)$. 
The RNN represents the function 
${\boldsymbol h}_t = f({\boldsymbol h}_{t-1},{\boldsymbol x}_t)$, 
where ${\boldsymbol h}_t \in \Re^{512}$ is a hidden state at position $t$ and
${\boldsymbol x}_t \in \Re^{64}$ is the one-hot coded vector 
of input symbol $s_t$.
The function $f$ is implemented by two stacked 
gated recurrent units (GRUs)~\cite{DBLP:conf/emnlp/ChoMGBBSB14}, 
each with 256 dimensional hidden states. 
The input vector ${\boldsymbol x}_t$ is fed to the lower GRU, 
and the hidden state of the lower GRU is fed to the upper GRU. 
The distribution of output symbol is computed as 
$P(y_t=j) = g_j({\boldsymbol h}_t)$, 
where $g_j$ is a softmax activation function 
depending only on the hidden state of the upper GRU. 

Given $N$ strings in the training set, 
we train the network such that it outputs a right-shifted version of the input.
Denote by ${\boldsymbol x}_{it}$ the one-hot coded vector of 
the $t$-th symbol in the $i$-th training string. 
The parameters in the network ${\boldsymbol \theta}$ 
is trained to minimize the following loss function,
\[
\min_{\boldsymbol \theta} \sum_{i=1}^N \sum_{t=1}^{T-1} D({\boldsymbol x}_{it+1}, P(y_t)),
\]
where $D$ denotes the relative entropy.
Our RNN was implemented using Keras library (\url{github.com/fchollet/keras}), 
and trained with ADAM~\cite{kingma2014adam} using a batch size of 256.
After the training is finished, one can compute $P(y_t)$ 
from $s_1,\ldots,s_{t-1}$. 
It allows us to perform rollout by sampling the next symbol repeatedly. 

\section{Experiments}
Following~\cite{gomez2016automatic}, we generate molecules that jointly 
optimize the octanol-water partition coefficient logP 
and the other two properties: 
synthetic accessibility~\cite{ertl2009estimation} and 
ring penalty that penalizes unrealistically large rings.
The score of molecule $S$ is described as
\begin{equation}\label{score}
J(S)=logP(S)-SA(S)-RingPenalty(S).
\end{equation}
The reward function of ChemTS is defined as 
\begin{equation} \label{eq:reward1}
r(S) = \left\{
\begin{array}{ll}
\frac{J(S)}{1+|J(S)|}  & \rm{Valid}\;\rm{SMILES} \\
-1.0  & \rm{otherwise.} \\
\end{array} \right.
\end{equation}
ChemTS was compared with two existing methods 
CVAE~\cite{gomez2016automatic} and GVAE~\cite{kusner2017grammar} 
based on variational autoencoders. 
Their implementation is available at \url{https://github.com/mkusner/grammarVAE}. 
Both methods perform molecular generation by Bayesian optimization (BO) in a latent space of VAE.
RNN, CVAE and GVAE were trained with 
approximately 250,000 molecules in ZINC database~\cite{irwin2012zinc}.
All methods were trained for 100 epochs.
Training took 3.8, 9.4 and 33.5 hours 
respectively, on a CentOS 6.7 server with a GeForce GTX Titan X GPU.
To evaluate the efficiency of MCTS, we prepared two alternative 
methods using RNN. 
One is simple random sampling using RNN, 
where the first symbol is made randomly and it is elongated until the terminal symbol occurs.
The other is the combination of RNN and Bayesian optimization~\cite{Ueno2016}, where 4,000 molecules are made a priori 
and Bayesian optimization is applied to find the best scoring molecule.

\begin{table}
\centering
\caption{Maximum score $J$ at time points 2,4,6 and 8 hours achieved by different molecular generation methods. The rightmost column shows the number of generated molecules per minute. The average values and standard deviations over 10 trials are shown.}
\label{tab:logp}
\resizebox{\textwidth}{!}{
\begin{tabular}{cccccc}
\hline
Method & 2h & 4h & 6h & 8h & Molecules/Min \\
\hline
ChemTS & $4.91\pm 0.38$ &  $5.41\pm 0.51$ & $5.49\pm 0.44$ & $5.58\pm 0.50$ & $40.89\pm 1.57$ \\
RNN+BO & $3.54\pm 0.27$ & $4.46\pm 0.24$ & $4.46\pm 0.24$ & $4.46\pm 0.24$ & $8.33\pm 0.00$ \\
Only RNN & $4.51\pm 0.27$ & $4.62\pm 0.26$ & $4.79\pm 0.25$ & $4.79\pm 0.25$ & $41.33\pm 1.42$ \\
CVAE+BO & $-30.18\pm 26.91$ & $-1.39\pm 2.24$ & $-0.61\pm 1.08$ & $-0.006\pm 0.92$ & $0.14\pm 0.08$ \\
GVAE+BO & $-4.34 \pm 3.14$ & $-1.29\pm 1.67$ & $-0.17\pm 0.96$ &  $0.25\pm 1.31$ &$1.38\pm 0.91$ \\
\hline
\end{tabular}}
\end{table}

\begin{figure*}
\resizebox{\textwidth}{!}{
\begin{tabular}{lllc}
	\hline
	&SMILES representation& $J$\\
	\hline
	
	     & 
\textcolor{blue}{O=C(Nc1cc(Nc2c(Cl)cccc2NCc2ccc(Cl)cc2Cl)c2}ccccc2c1OC(F)F)c1cccc2ccccc12 & 6.56\\

      &
\textcolor{blue}{O=C(Nc1cc(Nc2c(Cl)cccc2NCc2ccc(Cl)cc2Cl)ccc1C1=}CCCCC1)c1cc(F)cc(Cl)c1 & 6.43\\

	 & \textcolor{blue}{O=C(Nc1cc(Nc2c(Cl)cccc2N=C(SC}2CCCCC2)c2ccccc2)cc(Cl)c1Cl)c1ccc2ccccc2n1& 6.34\\

      & \textcolor{blue}{O=C(Nc1cc(Oc2ccc(Cl)cc2Cl)ccc1N}c1cc(Cl)ccc1Cl)c1ccc(Cl)cc1 & 6.33\\

      & 
      \textcolor{blue}{O=C(Nc1cc(Nc2c(Cl)cccc2Cl)c(Cl)cc1Br)N}(c1ccccc1)c1ccc(Cl)cc1 & 6.26\\

      & \textcolor{blue}{O=C(Nc1cc(Oc2c(Cl)cccc2Oc2ccc(-}c3ccccc3)cc2)ccc1Cl)c1ccccc1& 6.19\\

      & \textcolor{blue}{O=C(Nc1cc(Nc2c(Cl)cccc2Cl)c(Cl)c(C(=O)N(C}c2ccccc2)c2ccccc2)c1Cl)c1ccccc1F& 6.08\\

      & \textcolor{blue}{O=C(Nc1cc(Oc2ccc(Cl)cc2Cl)cc(Cl)c1Cl)c1nco}c1-c1ccc(Sc2ccccc2)cc1 & 6.007\\

      & \textcolor{blue}{O=C(Nc1cc(Nc2c(Cl)cccc2NCc2ccc(Cl)cc2Cl)c2ncccc2c1Cl)c1ccc(}Cl)cc1 & 6.0067\\

      &\textcolor{blue}{O=C(Nc1cc(Nc2c(Cl)cccc2NCc2ccc(Cl)cc2)c(}Cl)cc1Cl)c1cc(F)ccc1Cl& 6.0062\\

      &\textcolor{blue}{O=C(Nc1cc(Oc2c(Cl)cccc2Oc2ccccc2)nn}c1-c1ccccc1)c1sc2ccccc2c1Cc1ccccc1& 6.004\\

      & \textcolor{blue}{O=C(Nc1cc(Nc2c(Cl)cccc2NCc2ccc(Cl)cc2Cl)c2ncccc2c1Cl)c1ccccc1}Cl & 5.97\\

      & \textcolor{blue}{O=C(Nc1cc(Nc2c(Cl)cccc2NCc2ccc(Cl)cc2Cl)c(Cl)cc1Cl)c1ccc(}F)cc1F& 5.958\\

      &\textcolor{blue}{O=C(Nc1cc(Nc2c(Cl)cccc2NCc2ccccc2)ccc1C}(F)(F)F)c1ccc(Cl)c2ccccc12& 5.952\\

      &\textcolor{blue}{O=C(Nc1cc(Nc2c(Cl)cccc2Cl)c(Cl)cc1O}C(F)F)N(Cc1ccccc1)c1ccccc1C(F)(F)F & 5.94\\

      &\textcolor{blue}{O=C(Nc1cc(Oc2c(Cl)cccc2Oc2ccccc2C2=}CCCCC2)cc(Cl)c1)c1ccccc1& 5.93\\

      & \textcolor{blue}{O=C(Nc1cc(Nc2c(Cl)cccc2[N+]}(=O)[O-])cs1)c1sc2ccc(Br)cc2c1N(c1ccccc1)c1ccccc1& 5.92\\

      &\textcolor{blue}{O=C(Nc1cc(Nc2c(Cl)cccc2Cl)c(C}(=O)c2ccc(Cl)cc2F)c(Cl)c1)Nc1cccc(Cl)c1& 5.87\\
      &\textcolor{blue}{O=C(Nc1cc(Nc2c(Cl)cccc2NCc2ccc(Cl)cc2Cl)cc}(F)c1F)c1cccc2ccccc12 & 5.84\\

      & \textcolor{blue}{O=C(Nc1cc(Nc2c(Cl)cccc2NCc2ccc(Cl)cc2Cl)c(Cl)cc1Cl)c1cccs}1& 5.82\\
\hline
	\hline
\end{tabular}}

\vspace*{10mm}

  \includegraphics[width=\textwidth]{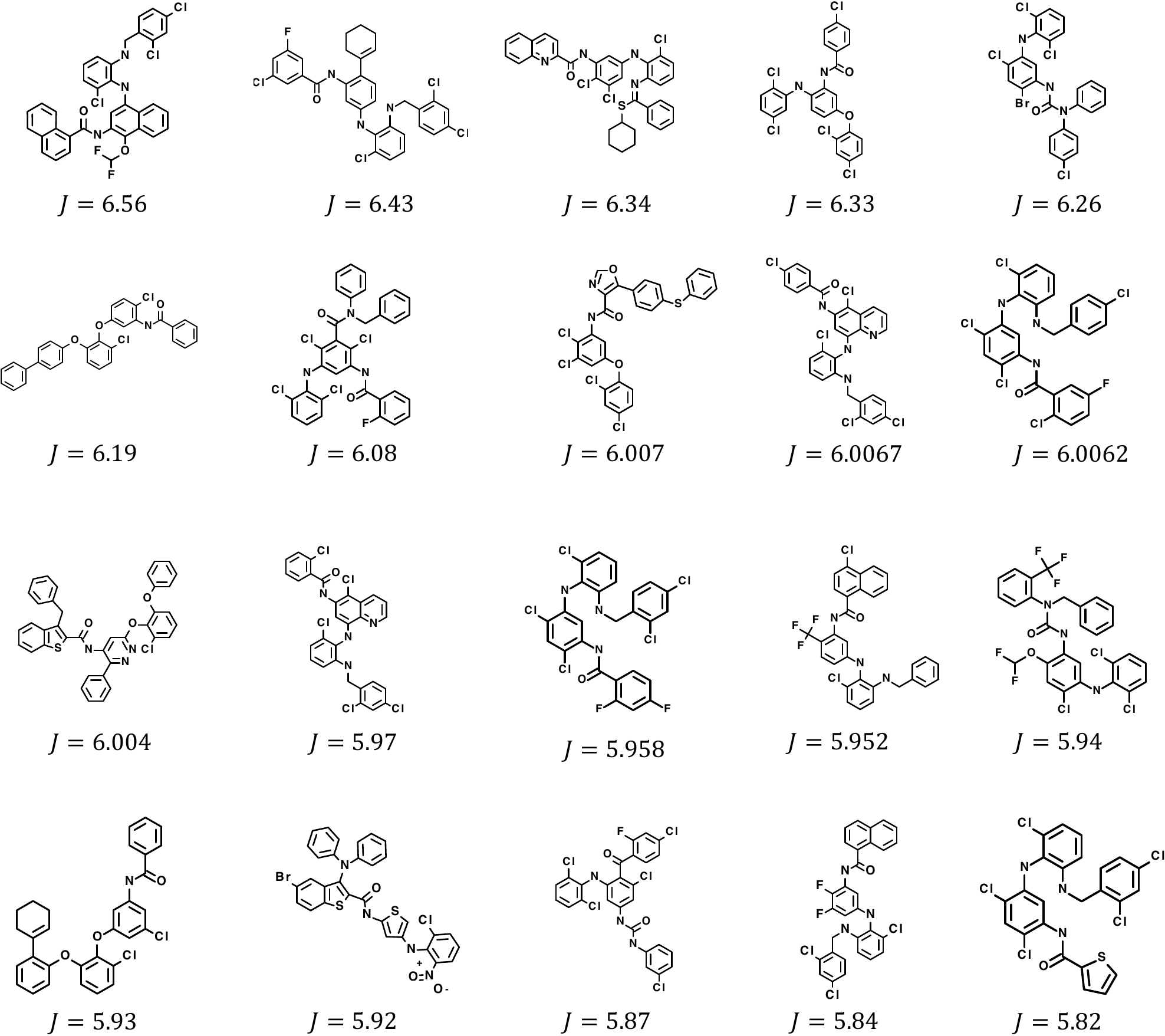}
  \caption{Best 20 molecules by ChemTS. Blue parts in SMILES strings indicate
  prefixes made in the search tree. The remaining parts are made by the
  rollout procedure.}
  \label{fig:best}
\end{figure*}

As shown in Table~\ref{tab:logp},
effectiveness of each method is quantified by the maximum score $J$ among all generated molecules at 2,4,6 and 8 hours  
and the speed of molecules generation (i.e., the number of generated
molecules per minute). 
VAE methods performed substantially slower than RNN-based methods, 
which reflects the low probability of generating valid SMILES strings.
ChemTS performed best in finding high scoring molecules, 
while the speed of molecular generation (40.89 molecules per minute)
was only slightly worse than
random generation by RNN (41.33 molecules per minute).
The combination of RNN and BO could not find high scoring molecules. 
Preparing more candidate molecules may improve the best score, but it
would further slow down the molecular generation. 
In general, it is difficult to design a correct reward function when
there are multiple objectives. 
So, it is important to generate many good molecules 
in a given time frame to allow the user to browse and select 
favorite molecules afterwards.
See Figure~\ref{fig:best} for the best molecules generated by ChemTS.

%\begin{figure*}
% \centering
%     \includegraphics[width=0.45\textwidth]{depthdis.pdf}
%     %\includegraphics[width=0.45\textwidth]{tani.pdf}
%     
%  \caption{Depth of the search tree in ChemTS. This distribution was obtained by one run of ChemTS with parameter $C=1.0$.}
%\label{fig:depth}
%\end{figure*}

\section{Conclusion}
In this paper, we presented a new python package for molecular
generation. It will be further extended to include more sophisticated
tree search methods and neural networks.
Use of additional packages for computational physics 
such as {\em pymatgen}~\cite{ong2013python} 
allows the users to implement their own reward function easily.
We look forward to see ChemTS as a part of the open-source ecosystem for
organic materials development.

\section*{Acknowledgement(s)}
We would like to thank Hou Zhufeng, Diptesh Das, Masato Sumita and Thaer M. Dieb for their fruitful discussions.
%We also would like to thank anonymous referees for their comments and suggestions to improve the manuscript.

\section*{Disclosure statement}
Authors declare no conflict of interest.

\section*{Funding}
This work was supported by the ``Materials research by Information Integration'' Initiative (MI2I) project and
CREST Grant No. JPMJCR1502 from Japan Science and Technology Agency (JST).
It was also supported by Grant-in-Aid for Scientific Research on
Innovative Areas ``Nano Informatics'' (Grant No. 25106005)
from the Japan Society for the Promotion of Science (JSPS).
In addition, it was supported by MEXT as ``Priority Issue on Post-K
computer'' (Building Innovative Drug Discovery Infrastructure Through Functional Control of
Biomolecular Systems).

\section*{Notes on contributors}

K. Tsuda proposed the research idea, supported experiments design 
and help draft the manuscript. 
K. Yoshizoe supported ChemTS design. 
J. Zhang and K. Terayama helped analyze the experimental data. 
X. Yang designed and implemented ChemTS, 
analyzed the data, and compiled the manuscript. 
All of the authors have read and approved the final manuscript.

%\bibliographystyle{tfnlm}
%\bibliography{interactnlmsample}

\end{document}